\documentclass[12pt]{article}
\usepackage{graphicx}
\usepackage{amsmath}
\usepackage{amssymb}
\usepackage{caption2}
\def\be{\begin{eqnarray}}
\def\en{\end{eqnarray}}
\def\non{\nonumber\\}

\def\prd{{Phys. Rev. D}~}

\def\plb{{ Phys. Lett. B}~}

\newcommand{\etap}{\eta^{\prime} }
\newcommand{\etapp}{\eta^{(\prime)}}

\newcommand{\psl}{ P \hspace{-2.4truemm}/ }
\newcommand{\nsl}{ n \hspace{-2.2truemm}/ }
\newcommand{\vsl}{ v \hspace{-2.2truemm}/ }
\newcommand{\epsl}{\epsilon \hspace{-1.8truemm}/\,  }
\begin{document}
\bibliographystyle{prsty}
\begin{center}
{\large {\bf {  The $B\to D_s^{(*)}\etapp$ Decays in the perturbative QCD  }}} \\[2mm]
Zhi-Qing
Zhang\footnote{Electronic address: zhangzhiqing@haut.edu.cn} \\
 Department of Physics, Henan University of Technology,
Zhengzhou, Henan 450001, P.R.China
\end{center}

\begin{abstract}
In this paper, we calculate the branching ratios for  $B^+\to
D_s^+\eta, B^+\to D_s^+\etap$, $B^+\to D_s^{*+}\eta$ and $ B^+\to
D_s^{*+}\etap$ decays by employing the perturbative QCD (pQCD)
factorization approach. Under the two kinds of $\eta-\etap$ mixing
schemes, the quark-flavor mixing scheme and the singlet-octet mixing
scheme, we find that the calculated branching ratios are consistent with
the currently available experimental upper limits. We also
considered the so called "$f_{D_s}$ puzzle", by using two groups of
parameters about the $D^{(*)}_s$ meson decay constants, that is
$f_{D_s}=241$ MeV, $f_{D^*_s}=272$ MeV and $f_{D_s}=274$ MeV,
$f_{D^*_s}=312$ MeV, to calculate the branching ratios for the considered
decays. We find that the results change $30\%$ by using these two different
groups of paramters.
\end{abstract}

PACS number:~ 13.25.Hw, 12.38.Bx, 14.40.Nd

Key words: B meson decay; the pQCD factorization approach; Branching ratio

\section{Introduction}\label{intro}
Recent years more and more efforts have been made to the B meson decays with one \cite{cdlv1} even two \cite{cdlv2}
charmed mesons in the final states and it is found that the perturbative QCD factorization (pQCD) approach do work well in these decays. So we are going to use
this approach to the decays involved one charmed meson $D^{(*)}_s$ and a light meson $\etapp$, which are shown in Fig.~1. The momenta of the two
 outgoing mesons are both approximately $\frac{1}{2}m_B(1-m^{2}_{D^{(*)}_s}/m_B^2)$. This is still large enough to make a hard intermediate gluon
in the hard part calculation.  Most of the momenta
come from the heavy $b$ quark in quark level. The light quark $u (d)$ inside $B^+$ $(B^0)$ meson, which is usually
called spectator quark, carries small momentum at order of $\Lambda_{QCD}$. In order to form a fast moving light meson, the
spectator quark need to connect the four quark operator through an energetic gluon. The hard four quark dynamic together with the
spectator quark becomes six-quark effective interaction. Since six-quark interaction is hard dynamics, it is perturbatively
calculable.

On the experimental side, the branching ratios of $B^+\to D_s^+
\etap$ and $B^+\to D_s^{*+}\etap$ decays have not been measured so for. For
$B^+\to D_s^+\eta$ and $B^+\to D_s^{*+}\eta$ decays, only the experimental limits are avaliable now\cite{pdg08}:
\be Br(B^+\to
D_s^+\eta) &<& 4.0\times 10^{-4}, \non Br(B^+\to D_s^{*+}\eta) &<&
6.0\times 10^{-4}.
 \en

In this paper, we will study the branching ratios of $B^+\to D_s^+\eta,
D_s^+\etap$ and $B^+\to D_s^{*+}\eta, D_s^{*+}\etap$ decays within perturbative
QCD approach based on $k_T$ factorization.
It is organized as follows. In Sec. \ref{proper}, the light-cone wave functions of the initial and
the final state mesons   are  discussed.
In Sec. \ref{results}, we then calculate analytically  these decay channels using the pQCD approach under the two
kinds of $\eta-\etap$ mixing schemes.
The numerical results and the discussions are given
in Sec. \ref{numer}. The conclusions are presented in the final part.

%=======================================================================
%       Physical properties of $f_0(980)$ and $f_0(1500)$
%=======================================================================

\section{Wave functions of initial and final state mesons}\label{proper}
In pQCD calculation, the light-cone wave functions of the mesons are nonperturbative part and not calculable in principal. But they
are universal and channel independent for all the hadronic decays. There are two heavy mesons in the each considered decay channels,
B and $D^{(*)}_s$.

In general, the B meson light-cone matrix element can be decomposed
as \cite{grozin}
\be
&&\int_0^1\frac{d^4z}{(2\pi)^4}e^{i\bf{k_1}\cdot z}
   \langle 0|\bar{b}_\alpha(0)d_\beta(z)|B(p_B)\rangle \nonumber\\
&=&-\frac{i}{\sqrt{2N_c}}\left\{(\psl_B+m_B)\gamma_5 \left[\phi_B
({\bf k_1})-\frac{\nsl-\vsl}{\sqrt{2}} \bar{\phi}_B({\bf
k_1})\right]\right\}_{\beta\alpha}, \label{aa1}
\en
where $n=(1,0,{\bf 0_T})$, and $v=(0,1,{\bf 0_T})$ are the
unit vectors pointing to the plus and minus directions,
respectively. Because the contribution of the second Lorentz structure $\bar{\phi}_B(x,b)$ is numerically small and can
be neglected. Therefore, we only consider the
contribution of Lorentz structure:
\be
\Phi_B(x,b)=
\frac{1}{\sqrt{2N_c}} (\psl_B +m_B) \gamma_5 \phi_B (x,b).
\label{bmeson}
\en
.

In heavy quark limit, we take the wave functions for the pseudoscalar meson $D_s$ and the vector meson $D^*_s$ as:
\be
\Phi_{D_S}(x,b)=
\frac{1}{\sqrt{2N_c}} \gamma_5(\psl_{D_s} +m_{D_s})  \phi_{D_s} (x,b),\\
\Phi_{D^*_s}(x,b)=
\frac{1}{\sqrt{2N_c}} \epsl(\psl_{D^*_s} +m_{D^*_s})  \phi_{D^*_s} (x,b),
\label{dsp}
\en
where the polar vector $\epsl=\frac{M_B}{\sqrt{2}M_{D^*_s}}(1,-r^2_{D^*_s},{\bf 0_T})$. In the considered decays,
$D^*_s$ meson is longitudinally polarized, so we only need consider its wave function
in longitudinal polarization.

The wave function for the effective quark  component $n\bar{n}$, which represents $u\bar{u}$ or $d\bar{d}$, of
$\eta^{(\prime)}$ meson is given as
\be \Phi_{\eta_{n\bar{n}}}(P,x,\zeta)\equiv
\frac{1}{\sqrt{2N_C}} \gamma_5 \left [ \psl
\phi_{\eta_{n\bar{n}}}^{A}(x)+m_0^{\eta_{n\bar{n}}}
\phi_{\eta_{n\bar{n}}}^{P}(x)+\zeta m_0^{\eta_{n\bar{n}}} ( \vsl
\nsl - v\cdot n)\phi_{\eta_{n\bar{n}}}^{T}(x) \right ], \label{etadis}
\en
where $P$ and $x$ are the momentum and the momentum fraction of
$\eta_{n\bar{n}}$, respectively.
The parameter $\zeta$ is either $+1$ or $-1$ depending on the
assignment of the momentum fraction $x$. For convenience, $
\phi_{\eta_{n\bar{n}}}^{A(P,T)}$ are denoted as $\phi_{\eta}^{A(P,T)}$ in the following.
The $s\bar s$ components of $\etapp$ are not relevant in these considered decays,
so we do not show their wave functions.
%===========================================================================
%                    Decay amplitudes in PQCD approach
%============================================================================

\section{ Perturbative QCD  calculation} \label{results}

Using factorization theorem, we can separate the decay amplitude into soft, hard, and harder dynamics characterized
by different scales, conceptually expressed as the convolution,
\be
{\cal A}(B \to D^{(*)}_s\etapp)\sim \int\!\! d^4k_1
d^4k_2 d^4k_3\ \mathrm{Tr} \left [ C(t) \Phi_B(k_1) \Phi_{D^{(*)}_s}(k_2)
\Phi_{\etapp}(k_3) H(k_1,k_2,k_3, t) \right ], \label{eq:con1}
\en
where $k_i$'s are momenta of the anti-quarks included in each mesons, and
$\mathrm{Tr}$ denotes the trace over Dirac and color indices. $C(t)$
is the Wilson coefficient which results from the radiative
corrections at short distance. In the above convolution, $C(t)$
includes the harder dynamics at larger scale than $M_B$ scale and
describes the evolution of local $4$-Fermi operators from $m_W$ (the
$W$ boson mass) down to $t\sim\mathcal{O}(\sqrt{\bar{\Lambda} M_B})$
scale, where $\bar{\Lambda}\equiv M_B -m_b$. The function
$H(k_1,k_2,k_3,t)$ describes the four quark operator and the
spectator quark connected by
 a hard gluon whose $q^2$ is in the order
of $\bar{\Lambda} M_B$, and includes the
$\mathcal{O}(\sqrt{\bar{\Lambda} M_B})$ hard dynamics. Therefore,
this hard part $H$ can be perturbatively calculated. The functions
$\Phi_{(D^{(*)}_s, \etapp)}$ are the wave functions of $D^{(*)}_s$ and $\etapp$, respectively.

In our paper, the light cone coordinate $(p^+,
p^-, {\bf p}_T)$ is used to describe the meson's momenta,
\be p^\pm =\frac{1}{\sqrt{2}} (p^0 \pm p^3), \quad {\rm and} \quad {\bf p}_T =
(p^1, p^2). \en
At the rest frame of $B$ meson, the light meson moves very fast and so $P_3^+$ or $P_3^-$ can be treated as zero.
The B meson and the two final state meson momenta can be written as
\be P_B =\frac{M_B}{\sqrt{2}} (1,1,{\bf 0}_T), \quad P_{2} =
\frac{M_B}{\sqrt{2}}(1,r^2,{\bf 0}_T), \quad P_{3} =
\frac{M_B}{\sqrt{2}} (0,1-r^2,{\bf 0}_T), \en
respectively, where $r=M_{D^{(*)}_s}/M_B$. Putting the anti-quark momenta in $B$,
$D^{(*)}_s$ and $\etapp$ mesons as $k_1$, $k_2$, and $k_3$, respectively, we can
choose
\be k_1 = (x_1 P_1^+,0,{\bf k}_{1T}), \quad k_2 = (x_2
P_2^+,0,{\bf k}_{2T}), \quad k_3 = (0, x_3 P_3^-,{\bf k}_{3T}). \en
For these considered decay channels, the integration over $k_1^-$,
$k_2^-$, and $k_3^+$ in eq.(\ref{eq:con1}) will lead to
\be
 {\cal
A}(B \to D^{(*)}_s\etapp ) &\sim &\int\!\! d x_1 d x_2 d x_3 b_1 d b_1 b_2 d
b_2 b_3 d b_3 \non && \cdot \mathrm{Tr} \left [ C(t) \Phi_B(x_1,b_1)
\Phi_{D^{(*)}_s}(x_2,b_2) \Phi_{\etapp}(x_3, b_3) H(x_i, b_i, t) S_t(x_i)\,
e^{-S(t)} \right ], \quad\;\;\;\label{eq:a2}
\en
where $b_i$ is the
conjugate space coordinate of $k_{iT}$, and $t$ is the largest
energy scale in function $H(x_i,b_i,t)$. The
last term $e^{-S(t)}$ in Eq.(\ref{eq:a2}) is the Sudakov form factor which suppresses
the soft dynamics effectively \cite{soft}.

\begin{figure}
 \centering\vspace{-3cm}
 \includegraphics[totalheight=22cm,width=15cm]{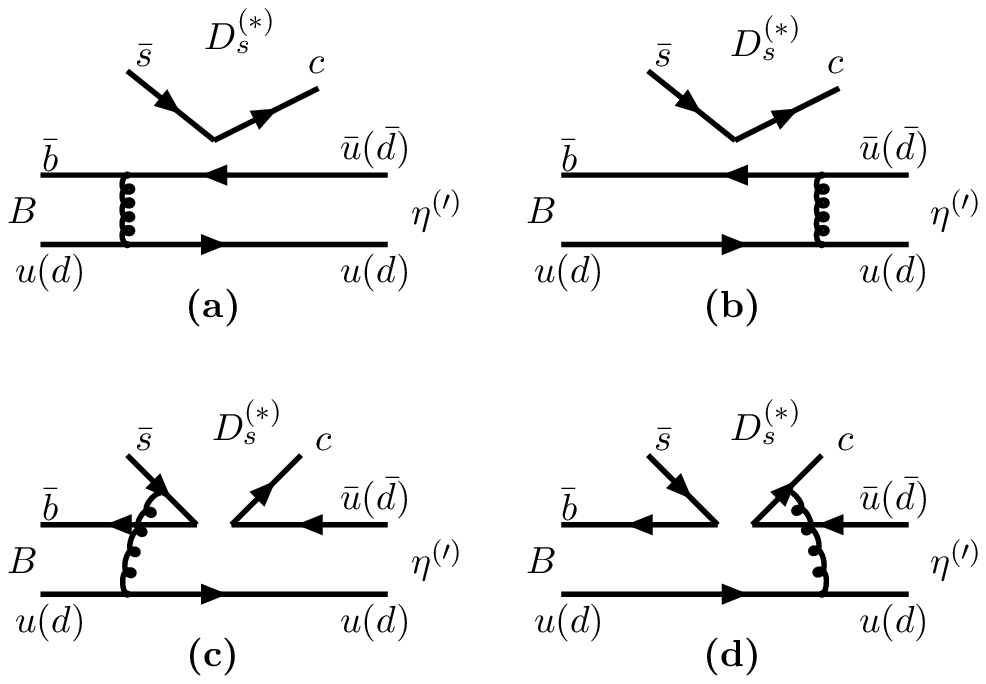}
\vspace{-14cm}
\caption{ Diagrams contributing to the decays $B\to D^{(*)}_s\etapp$.}
 \label{fig1}
\end{figure}

 For the considered decays, the related weak effective
Hamiltonian $H_{eff}$ can be written as \cite{buras96}
\be
\label{eq:heff} {\cal H}_{eff} = \frac{G_{F}} {\sqrt{2}} \,
V_{ub}^* V_{cs}\left[ \left (C_1(\mu) O_1(\mu) +
C_2(\mu) O_2(\mu) \right)\right] ,\non
\;\;\;\;\;O_1=(\bar b_\alpha u_\beta)_{V-A}(\bar c_\alpha s_\beta)_{V-A}  ,\quad O_2  = (\bar b_\alpha u_\alpha)_{V-A}(\bar c_\alpha s_\alpha)_{V-A} ,
\en
where the Fermi constant $G_{F}=1.166 39\times
10^{-5} GeV^{-2}$, the CKM matrix elements $|V_{ub}|=(3.93\pm0.36)\times10^{-3}, |V_{cs}|=1.04\pm0.06$ \cite{pdg08},
$C_{1,2}(\mu)$ are Wilson coefficients running with the renormalization scale $\mu$. Here
$\alpha, \beta$ are the color indexes, and $(\bar{q}_1q_2)_{V-A}=\bar q_1\gamma^\mu(1-\gamma^5)q_2$.
The leading order diagrams contributing to the decays $B\to D^{(*)}_s\etapp$ are drawn in Fig.\ref{fig1}
according to this effective Hamiltonian.

%===========================================================================
%                  Numerical results and discussions
%============================================================================

In the following, we take the $B\to D^{(*)}_s\eta$ decay
channel as  an example to get the analytic formulas by calculating the hard part $H(t)$ at leading order.
Involving the meson wave functions, the amplitude for the factorizable tree emission diagrams Fig.1(a) and (b)
can be written as:
\be
F_{e\eta}&=&8\pi C_Ff_{D^{(*)}_s}\int_0^1 dx_1 dx_3 \int_0^{\infty} b_1db_1\, b_3db_3\,
\Phi_B(x_1,b_1)\nonumber \\
& &\times
\left\{\left[(x_3+1)\phi^A_{\eta}(x_3)-r_\eta(2x_3-1)(\phi^P_{\eta}(x_3)+\phi^T_{\eta}(x_3))
\right]\right.\non &&\left.
\times E_e(t)h_e(x_1,x_3(1-r^2_{D^{(*)}_s}),b_1,b_3)S_t(x_3)\right.\non &&\left.
+2r_\eta\phi^P_{\eta}(x_3)E_e(t')h_e(x_3,x_1(1-r^2_{D^{(*)}_s}),b_3,b_1)S_t(x_1)\right\}\;, \label{fe}
\en
where $C_F=4/3$ is the group factor of $SU(3)_c$ gauge group, and the mass ratios $r_\eta=m_0^{\eta_{n\bar{n}}}/m_B, r_{D^{(*)}_s}
=m_{D^{(*)}_s}/m_B$ and $f_{D^{(*)}_s}$ is the decay constant of $D^{(*)}_s$ meson. The factor evolving with the scale
$t$ is given by:
\be
E_e(t)=\alpha_s(t)\exp[-S_B(t)-S_{\eta}(t)],
\en
where the expressions for Sudakov form factors $S_B(t),S_{\eta}(t)$ and the jet function $S_t(x)$ can be found
in \cite{PQCD}. The hard function is written as:
\be
h_e(x_1,x_2,b_1,b_2)&=&  K_{0}\left(\sqrt{x_1 x_2} m_B b_1\right)
\left[\theta(b_1-b_2)K_0\left(\sqrt{x_2} m_B
b_1\right)I_0\left(\sqrt{x_2} m_B b_2\right)\right. \non & &\;\left.
+\theta(b_2-b_1)K_0\left(\sqrt{x_2}  m_B b_2\right)
I_0\left(\sqrt{x_2}  m_B b_1\right)\right]. \label{he1}
\en
The hard scales $t$ and $t'$ in Eq.(\ref{fe}) are determined by
\be
t&=&max(\sqrt{x_3(1-r^2_{D^{(*)}_s})}m_B,1/b_1,1/b_3),\non  t'&=&max(\sqrt{x_1(1-r^2_{D^{(*)}_s})}m_B,1/b_1,1/b_3).
\label{scale1}
\en

For the nonfactorizable tree emission diagrams Fig.1(c) and (d), all three meson wave functions are
involved. The integraton of $b_3$ can be performed using $\delta$ function $\delta(b_3-b_2)$ and the result is
\be
M_{e\eta}&=&-16\pi\sqrt{2N_c}C_F\int_0^1 dx_1dx_2dx_3 \int_0^{\infty} b_1 db_1\, b_2
db_2\,\Phi_B(x_1,b_1) \Phi_{D^{(*)}_s}(x_2) \non
&&\times \left\{\left[(x_2-1)\phi^A_{\eta}(x_3)+r_\eta x_3(\phi^P_{\eta}(x_3)-\phi^T_{\eta}(x_3))\right]
E_n(t)h^1_n(x_1,x_2,x_3,b_1,b_2)\right.\non &&\left.
+[(x_3+x_2)\phi^A_{\eta}(x_3)-r_\eta x_3(\phi^P_{\eta}(x_3)+\phi^T_{\eta}(x_3))E_n(t')h^2_n(x_1,x_2,x_3,b_1,b_2)]\right\},
\en
where the expression for the evolution factor is $E_n=\alpha_s(t)\exp[-S(t)|_{b_3=b_1}]$ with the Sudakov exponent
$S=S_B+S_{D^{(*)}_s}+S_\eta$.

The hard functions $h^{i}_{n}, i=1,2$ in the amplitude are given as
\be
h^i_n&=&[\theta(b_1-b_2)K_0(Ab_1)I_0(Ab_2)+\theta(b_2-b_1)K_0(Ab_2)I_0(Ab_1)]\non &&
\non && \times\left(
\begin{matrix}
 \frac{\pi i}{2}\mathrm{H}_0(\sqrt{|G^2_i|} b_3), & \text{for}\quad G^2_i<0 \\
 \mathrm{K}_0(G_ib_2), &
 \text{for} \quad G^2_i>0
\end{matrix}\right),\label{cdhard}
\en
with the variables
\be
A^2&=&x_1x_3(1-r^2_{D^{(*)}_s})m^2_B,\non
G^2_1&=&(x_1+x_2)r^2_{D^{(*)}_s}-(1-x_1-x_2)x_3(1-r^2_{D^{(*)}_s})m^2_B,\non
G^2_2&=&(x_1-x_2)x_3(1-r^2_{D^{(*)}_s})m^2_B.
\en
The hard scales in Eq.(\ref{cdhard}) are given by
\be
t&=&max(Am_B,\sqrt{G^2_1}m_B, 1/b_1, 1/b_2), \non t'&=&max(Am_B,\sqrt{G^2_2}m_B, 1/b_1, 1/b_2).\label{scale}
\en

Before we write down the complete decay amplitudes for the considered decay channels, we firstly give a brief
discussion about $\eta-\etap$ mixing. As it is well-known, there exist two popular mixing schemes,
quark-flavor mixing scheme (S1) and singlet-octet mixing scheme (S2) \cite{ekou,feld}. In the former the
$n\bar{n}=\frac{(u\bar u +d\bar d)}{\sqrt{2}}$ and $s\bar s$ flavor states, labeled by the $\eta_q$ and $\eta_s$
mesons, are defined. The physical states $\eta$
and $\eta^{\prime}$ are related to the flavor states through a
single angle $\phi$,
 \be \left(\begin{array}{c}
     \eta \\ \eta^{\prime} \end{array} \right)
=\left(\begin{array}{cc}
 \cos{\phi} & -\sin{\phi} \\
 \sin{\phi} & \cos{\phi} \\ \end{array} \right)
 \left(\begin{array}{c} \eta_q \\ \eta_s \end{array} \right).
\label{eq:e-ep} \en
In the latter the singlet-octet states $(u\bar u+d\bar d+s\bar s)/\sqrt{3}$ and $(u\bar u+d\bar d-2s\bar s)/\sqrt{6}$ ,
labeled by the
$\eta_1$ and $\eta_8$ mesons, are considered. The physical states $\eta$
and $\eta^{\prime}$ are related to the singlet-octet states through an
angle $\theta_p$,
 \be \left(\begin{array}{c}
     \eta \\ \eta^{\prime} \end{array} \right)
=\left(\begin{array}{cc}
 \cos{\theta_p} & -\sin{\theta_p} \\
 \sin{\theta_p} & \cos{\theta_p} \\ \end{array} \right)
 \left(\begin{array}{c} \eta_8 \\ \eta_1 \end{array} \right).
\label{singlet} \en
The mixing angle $\phi$ has been well determined, $\phi=39.3^\circ\pm1.0^\circ$. While one finds that the angle $\theta_p$ is in the
range of $-17^\circ\leq \theta_p\leq -10^\circ$ through fitting the various related experimental results \cite{ekou}.

Then the total decay amplitudes of $B\to D^{(*)}_s\eta$ channels can be written as
\be
{\cal A}(B\to D^{(*)}_s\eta)&=&
V_{ub}^* V_{cs}[F_{e\eta}(C_2+\frac{C_1}{3})+M_{e\eta}C_1]F_1(\phi) \quad \text{for S1}, \non
{\cal A}(B\to D^{(*)}_s\eta)&=&
V_{ub}^* V_{cs}[F_{e\eta}(C_2+\frac{C_1}{3})+M_{e\eta}C_1]F_1(\theta_p) \quad \text{for S2}
, \label{totalam}
\en
where
\be
F_1(\phi)=\frac{\cos{\phi}}{\sqrt{2}},\quad\quad
F_1(\theta_p)=-\frac{\sin{\theta_p}}{\sqrt{3}}+
\frac{\cos{\theta_p}}{\sqrt{6}},
\en
are the mixing factors in the quark-flavor and singlet-octet mixing schemes,
respectively. The decay amplitudes for $B\to D^{(*)}_s\eta'$ can be obtained easily from Eq.(\ref{totalam})
by the following replacements:
\be
\eta &\longrightarrow& \eta', \\
F_1(\phi) &\longrightarrow & F'_1(\phi) =\frac{1}{\sqrt{2}}\sin\phi,
\\ F_1(\theta_p) &\longrightarrow & F'_1(\theta_p) =\frac{\cos{\theta_p}}{\sqrt{3}}+\frac{\sin{\theta_p}}
{\sqrt{6}}. \en
\section{Numerical results and discussions} \label{numer}
For the $B$ meson wave function, we adopt the model
\be
\phi_B(x,b)
&=& N_B x^2(1-x)^2 \mathrm{exp} \left
 [ -\frac{M_B^2\ x^2}{2 \omega_{b}^2} -\frac{1}{2} (\omega_{b} b)^2\right],
 \label{phib}
\en
where $\omega_{b}$ is a free parameter and we take
$\omega_{b}=0.4\pm 0.04$ GeV in numerical calculations, and
$N_B=91.745$ is the normalization factor for $\omega_{b}=0.4$ GeV and $f_B = 0.19$ GeV.

The model of $\phi_{D^{(*)}_s}$ is adopted as
\be
\phi_{D^{(*)}_s}=f_{D^{(*)}_s}\frac{1}{\sqrt 6}x(1-x)\left[1-a_{D^{(*)}_s}(1-2x)\right], \label{phids}
\en
where
\be
a_{D_s}&=&0.3,\quad a_{D^{*}_s}=0.78, \non
f_{D_s}&=&0.241 \text{GeV}  , \quad f_{D^*_s}=0.272 \text{GeV} .\label{fdss}
\en
The values of these parameters can be found in the Refs.\cite{cdlv1,yuming,aubin,parodi,runhui}.

For $\eta$ meson's wave function, the distribution amplitudes
$\phi_{\eta}^A$, $\phi_{\eta}^P$ and
$\phi_{\eta}^T$ represent the axial vector, pseudoscalar
and tensor components of the wave function, respectively. They are given as:
\begin{eqnarray}
 \phi_{\eta}^A(x) &=&  \frac{f_{\eta_{n\bar{n}}}}{2\sqrt{2N_c} }
    6x (1-x)
    \left[1+a^{\eta_{n\bar{n}}}_1C^{3/2}_1(2x-1)+a^{\eta_{n\bar{n}}}_2C^{3/2}_2(2x-1)
    \right.\non && \left.+a^{\eta_{n\bar{n}}}_4C^{3/2}_4(2x-1)
  \right],\label{piw1}\\
 \phi_{\eta}^P(x) &=&   \frac{f_{\eta_{n\bar{n}}}}{2\sqrt{2N_c} }
   \left[ 1+(30\eta_3-\frac{5}{2}\rho^2_{\eta_{n\bar{n}}})C^{1/2}_2(2x-1)-3\left\{\eta_3\omega_3+\frac{9}{20}\rho^2_{\eta_{n\bar{n}}}
   (1+6a^{\eta_{n\bar{n}}}_2)\right\}\right. \non && \left.\times C^{1/2}_4(2x-1)\right]  ,\\
 \phi_{\eta}^T(x) &=&  \frac{f_{\eta_{n\bar{n}}}}{2\sqrt{2N_c} } (1-2x)
   \left[ 1+6(5\eta_3-\frac{1}{2}\eta_3\omega_3-\frac{7}{20}\rho^2_{\eta_{n\bar{n}}}-\frac{3}{5}\rho^2_{\eta_{n\bar{n}}}a_2^{\eta_{n\bar{n}}})
   (1-10x+10x^2)\right] ,\quad\quad\label{piw}
 \end{eqnarray}
with \be f_{\eta_{n\bar{n}}}&=&1.07f_\pi,\quad
\rho_{\eta_{n\bar{n}}}=2m_n/m_{nn},\non
a^{\eta_{n\bar{n}}}_1&=&0,\quad
a^{\eta_{n\bar{n}}}_2=0.115\pm0.115,\quad
a^{\eta_{n\bar{n}}}_4=-0.015. \label{fetann}\en
The chiral enhancement scale $m_0^{\eta_{n\bar{n}}}$ shown in Eq.(\ref{etadis}) is defined by
\be
m_0^{\eta_{n\bar{n}}}&=&\frac{m^2_{nn}}{2m_n}=\frac{1}{2m_n}[m^2_{\eta}\cos^2\phi
 +m^2_{\eta^{\prime}}\sin^2\phi-\frac{\sqrt{2}f_{\eta_{s\bar{s}}}}{f_{\eta_{n\bar{n}}}}(m^2_{\eta^{\prime}}-m^2_{\eta})
 \cos\phi\sin\phi],\label{eq:m0q}
\en
where the current quark mass $m_n=m_u=m_d$ and $f_{\eta_{s\bar{s}}}=(1.34\pm 0.06)f_{\pi}$ \cite{feld,xiao}.

The parameters defined in Eq.(\ref{fetann}) and Eq.(\ref{eq:m0q}) are for the quark-flavor mixing scheme.
As for singlet-octet mixing scheme, the parameters $f_{\eta_{n\bar{n}}}, m_0^{\eta_{n\bar{n}}}$ are changed to
$f_{\pi}$ and $m_0^\pi$, respectively.

In the B-rest frame, the decay rates of $B\to D^{(*)}_s\etapp$ can be written as:
\be
\Gamma=\frac{1}{32\pi}G_F^2m^7_B|{\cal A}|^2(1-r^2_{D^{(*)}_s})\;,
\en
where ${\cal A}$ is the total decay amplitude shown in Eq.(\ref{totalam}).

%%%%%%%%%%%%%%%%%%%%%%%%%%%%%%%%%%
\begin{table}
\caption{Branching ratios ($\times 10^{-5}$) for the decays $B^+\to D_s^{(*)+}\eta$ and
$B^+\to D_s^{(*)+}\etap$. S1 represents the quark-flavor mixing scheme with mixing angle $\phi=39.3^\circ$.
S2 represents the singlet-octet mixing scheme with mixing angle $\theta_p=-10^\circ$ and $\theta_p=-17^\circ$,
respectively. Here the threshold resummation parameter $c=0.3$.
The first theoretical error is from the B meson shape parameter $\omega_b$, the second error arises from the
higher order pQCD correction. The third one is from the uncertainties of CKM matrix elements.
We also list the current experimental upper limits ($90\%$ C.L.).}\label{brch1}
\begin{center}
\begin{tabular}{c|c|c|c|c}
   \hline \hline
   Channel & S1($\phi=39.3^\circ$) & S2($\theta_p=-10^\circ$) & S2($\theta_p=-17^\circ$) &Data  \\
   \hline
    $B^+\to D_s^+\eta $ &$1.08^{+0.36+0.30+0.21}_{-0.61-0.06-0.21}$& $0.80^{+0.26+0.27+0.15}_{-0.19-0.06-0.15}$
    &$0.99^{+0.34+0.29+0.18}_{-0.24-0.07-0.18}$
    &$<40$\\
    $B^+\to D_s^+\etap $&$0.73^{+0.24+0.23+0.14}_{-0.17-0.05-0.14}$ &$0.78^{+0.25+0.24+0.14}_{-0.19-0.05-0.14}$
    &$0.59^{+0.18+0.16+0.08}_{-0.10-0.03-0.08}$&--\\
    $B^+\to D_s^{*+}\eta$ &$1.44^{+0.48+0.35+0.27}_{-0.34-0.10-0.27}$&$1.06^{+0.34+0.29+0.20}_{-0.58-0.06-0.20}$
    &$1.31^{+0.43+0.34+0.25}_{-0.76-0.07-0.25}$&$<60$\\
    $B^+\to D_s^{*+}\etap$ &$0.97^{+0.32+0.29+0.18}_{-0.23-0.07-0.18}$&$1.04^{+0.33+0.29+0.19}_{-0.57-0.05-0.19}$
    &$0.79^{+0.26+0.26+0.14}_{-0.20-0.05-0.14}$&--\\
   \hline\hline
\end{tabular}
   \end{center}
\end{table}
%%%%%%%%%%%%%%%%%%%%%%%%%%%%%%%%%%%%%%%%%%

Using the wave functions and the input parameters as specified in the previous part, it is straightforward to calculate the
CP-averaged branching ratios for the considered decays, which are listed in Table~\ref{brch1}.
The first error in these entries is caused by the B meson shape parameter $\omega_b=0.40\pm0.04$.
The second error arises from the higher order pQCD correction: the choice of hard scales, which have
been defined in Eq.(\ref{scale1}) and Eq.(\ref{scale}), vary from $0.9t$ to $1.1t$. The third error is from the
uncertainties of the CKM matrix elements.

It is known that there is a discrepancy on the decay constant $f_{D_s}$ between theory and experiment \cite{zhengtao}, which is so called
"$f_{D_s}$ puzzle". For example, the
result within convariant light-front approach is about $230$ MeV \cite{haiyang}, and it is $f_{D_s}=249\pm3\pm16$ MeV given by
the Lattice QCD calculation \cite{aubin}. While the measurements of $f_{D_s}$ have been  improved by the CLEO and BarBar
collaborations \cite{cleo,barbar},
and obtained $274\pm13\pm7$ MeV and $283\pm17\pm7\pm14$ MeV, respectively. However, the decay constant of $D^{*}_s$
has not been directly measured in experiment so far. On the theoretical side, the result from the Lattice QCD
calculations shows $f_{D^*_s}=272\pm16^{+3}_{-20}$ MeV \cite{aubin,lattice,ukqcd}, which is used in our numerical  calculation. Certainly,
the consistent result is also obtained by the QCD sum rules, $f_{D^*_s}=260^{+9}_{-12}$ MeV \cite{yuming}.

If we use the decay constants given by theory shown in Eq.(\ref{fdss}), the branching ratios are listed in Table~\ref{brch1}.
We also calculate by using the improving measured value obtained by the CLEO
collaboration, $f_{D_s}=0.274$ GeV, at the same time, the decay constant of $D^*_s$ is taken as $f_{D^*_s}=0.312$ GeV
\cite{runhui1}. Then the corresponding results are listed in Table~\ref{brch2}. One can find
the branching ratios obtained a $30\%$ enhancement by using these new decay constants.

%%%%%%%%%%%%%%%%%%%%%%%%%%%%%%%%%%
\begin{table}
\caption{Branching ratios ($\times 10^{-5}$) for the decays $B^+\to D_s^{(*)+}\eta$ and
$B^+\to D_s^{(*)+}\etap$
with the decay constants $f_{D_s}=274$ MeV and $f_{D^*_s}=312$ MeV. Here the threshold resummation parameter $c=0.3$.
The errors for these entries correspond to the uncertanties of  the B meson shape parameter $\omega_b$,
 from the scale-dependence and  the CKM matrix elements, respectively.}\label{brch2}
\begin{center}
\begin{tabular}{c|c|c|c|c}
   \hline \hline
   Channel & S1($\phi=39.3^\circ$) & S2($\theta_p=-10^\circ$) & S2($\theta_p=-17^\circ$) &Data  \\
   \hline
    $B^+\to D_s^+\eta $ &$1.40^{+0.46+0.10+0.27}_{-0.38-0.12-0.27}$& $1.03^{+0.26+0.09+0.20}_{-0.30-0.07-0.20}$
    &$1.28^{+0.40+0.09+0.25}_{-0.32-0.10-0.25}$&$<40$\\
    $B^+\to D_s^+\etap $&$0.94^{+0.22+0.08+0.18}_{-0.28-0.07-0.18}$ & $1.01^{+0.25+0.09+0.19}_{-0.29-0.07-0.19}$
    &$0.77^{+0.26+0.23+0.14}_{-0.18-0.06-0.14}$&--\\
    $B^+\to D_s^{*+}\eta$ &$1.90^{+0.62+0.16+0.33}_{-0.45-0.13-0.33}$&$1.40^{+0.45+0.10+0.26}_{-0.38-0.12-0.26}$
    &$1.73^{+0.52+0.14+0.30}_{-0.40-0.15-0.30}$&$<60$\\
    $B^+\to D_s^{*+}\etap$ &$1.27^{+0.42+0.13+0.24}_{-0.30-0.09-0.24}$&$1.37^{+0.44+0.10+0.25}_{-0.36-0.12-0.25}$
    &$1.04^{+0.26+0.09+0.20}_{-0.28-0.07-0.20}$&--\\
   \hline\hline
\end{tabular}
   \end{center}
\end{table}

It is noticed that in the upper two groups of values for the decay constants $f_{D_s}$ and $f_{D^{*}_s}$, the relation \cite{yuming,ukqcd}
\be
\frac{f_{D^*_s}}{f_{D^*}}\approx\frac{f_{D_s}}{f_{D}}\approx\frac{f_{B_s}}{f_B}=[1.1,1.2]
\en
is connotative. It is different from \cite{runhui},
where the relation between $f_{D^*_s}$ and $f_{D_s}$ derived from HQET was used:
\be
\frac{f_{D^*_s}}{f_{D_s}}=\sqrt{\frac{m_{D_s}}{m_{D^*_s}}}\;. \label{hqet1}
\en
From the Eq.~(\ref{hqet1}), one can get the value of $f_{D^*_s}$, which is less than $f_{D_s}$.

%%%%%%%%%%%%%%%%%%%%%%%%%%%%%%%%%%%%%%%%
\begin{table}
\caption{Branching ratios ($\times 10^{-5}$) for the decays $B^+\to D_s^{(*)+}\eta$ and
$B^+\to D_s^{(*)+}\etap$. When we take the parameters in S1, the branching ratios for different threshold
resummation parameter $c$. The errors for these entries correspond to the uncertanties of  the B meson shape parameter $\omega_b$,
 from the scale-dependence and  the CKM matrix elements, respectively. }\label{brch3}
\begin{center}
\begin{tabular}{c|c|c|c}
   \hline \hline
   Channel & $c=0.3$ & $c=0.35$ & $c=0.4$   \\
   \hline
    $B^+\to D_s^+\eta $ &$1.08^{+0.36+0.30+0.21}_{-0.61-0.06-0.21}$& $0.97^{+0.31+0.29+0.17}_{-0.24-0.09-0.17}$
    &$0.85^{+0.28+0.26+0.15}_{-0.20-0.06-0.15}$\\
    $B^+\to D_s^+\etap $&$0.73^{+0.24+0.23+0.13}_{-0.17-0.05-0.13}$ &$0.65^{+0.21+0.21+0.12}_{-0.13-0.03-0.12}$
    &$0.57^{+0.19+0.18+0.10}_{-0.11-0.02-0.10}$\\
    $B^+\to D_s^{*+}\eta$ &$1.44^{+0.48+0.35+0.27}_{-0.84-0.10-0.27}$&$1.29^{+0.42+0.33+0.25}_{-0.76-0.08-0.25}$
    &$1.14^{+0.40+0.31+0.23}_{-0.70-0.07-0.23}$\\
    $B^+\to D_s^{*+}\etap$ &$0.97^{+0.32+0.29+0.18}_{-0.23-0.07-0.18}$&$0.86^{+0.28+0.27+0.16}_{-0.22-0.07-0.16}$&
    $0.76^{+0.25+0.23+0.13}_{-0.18-0.06-0.13}$\\
   \hline\hline
\end{tabular}
   \end{center}
\end{table}
%%%%%%%%%%%%%%%%%%%%%%%%%%%%%%%%%%%%%5%%
From Table~\ref{brch1} and Table~\ref{brch2}, one can find that the differences of
the branching ratios between $B^+\to D_s^{(*)+}\eta$ and $B^+\to D_s^{(*)+}\etap$ are small in S2 with
mixing angle $\theta_p=-10^\circ$, while they are large in S2 with mixing angle $\theta_p=-17^\circ$ and S1.
For the decays $B^+\to D_s^{(*)+}\eta$, their branching ratios are close in S1 and S2 with mixing angle
$\theta_p=-17^\circ$, respectively. For the decays $B^+\to D_s^{(*)+}\etap$, their branching ratios are close
in S1 and S2 with mixing angle $\theta_p=-10^\circ$.

If we take the different threshold resummation parameter c,
one can find the results (shown in Table~\ref{brch3}) corresponding to these
parameters $c=0.3, 0.35, 0.4$ decrease sequentially by about $10\%$.
It is noticed that the threshold resummation factor
is only considered in the factorizable contributions. It is the same with the decays $B\to  D_s^{(*)}\pi$, the
contributions from the factorizable amplitudes are dominant.

%===========================================================================
%                 Conclusion
%============================================================================

\section{Conclusion}\label{summary}

In this paper, we calculate the branching ratios of  decays $B^+\to D_s^+\etapp$ and $B^+\to D_s^{*+}\etapp$
in the pQCD factorization approach. We find that:
\begin{itemize}
\item
To determine the vector meson $D^{*}_s$ decay constant, the relation
$
\frac{f_{D^*_s}}{f_{D^*}}\approx\frac{f_{D_s}}{f_{D}}\approx\frac{f_{B_s}}{f_B}
$
is connotative. It implies that $f_{D^*_s}$ is a little larger than $f_{D_s}$, which is not the same with the
relation derived from HQET:
$
\frac{f_{D^*_s}}{f_{D_s}}=\sqrt{\frac{m_{D_s}}{m_{D^*_s}}}\;.
$
\item
For the branching ratios of the considered decay modes, the pQCD predictions in the quark-flavor mixing scheme are
\be
Br(B^+\to D_s^+\eta)=(1.08^{+0.36+0.30+0.21}_{-0.61-0.06-0.21})\times 10^{-5},\non
Br(B^+\to D_s^+\etap)=(0.73^{+0.24+0.23+0.14}_{-0.17-0.05-0.14})\times 10^{-5},\non
Br(B^+\to D_s^{*+}\eta)=(1.44^{+0.48+0.35+0.27}_{-0.34-0.10-0.27})\times 10^{-5},\non
Br(B^+\to D_s^{*+}\etap)=(0.97^{+0.32+0.29+0.18}_{-0.23-0.07-0.18})\times 10^{-5}.
\en .
\item
If we take the improved measurement $f_{D_s}=0.274$ GeV, which is obtained by the CLEO collaboration, and
$f_{D^*_s}=0.312$ GeV, the branching ratios get a $30\%$ enhancement.
\item
We also investigate the effect of the threshold resummation. Taking the different threshold parameter
$c=0.3, 0.35, 0.4$, one can find the results decrease sequentially by about $10\%$.

\end{itemize}

\section*{Acknowledgment}
Z.Q.~Zhang would like to thank C.D.~L\"u for fruitful discussions.

%%%%%%%%%%%%%%%%%%%%%%%%%%%%%%%%%%%%%%%%%%%%%%%%%%%%%%%%%%%%%%%%%%%%%%%%
%                               references
%%%%%%%%%%%%%%%%%%%%%%%%%%%%%%%%%%%%%%%%%%%%%%%%%%%%%%%%%%%%%%%%%%%%%%%%

\end{document}